\documentclass[conference, compsoc]{IEEEtran}
\usepackage{amsmath,amssymb,amsfonts}
\usepackage{booktabs}
\usepackage{spverbatim}
\usepackage{graphicx}
\usepackage{hyperref}
\usepackage{xurl}

\usepackage{refcount}
\newcounter{fncntr}
\newcommand{\fnmark}[1]{\refstepcounter{fncntr}\label{#1}\footnotemark[\getrefnumber{#1}]}
\newcommand{\fntext}[2]{\footnotetext[\getrefnumber{#1}]{#2}}

\begin {document}
\title{How segmented is my network?}

\author{\IEEEauthorblockN{Rohit Dube}
\IEEEauthorblockA{\textit{Independent Researcher} \\
                  California, USA \\
                 }
}

\maketitle

\begin{abstract}
Network segmentation is a popular security practice for limiting lateral movement, yet practitioners lack a metric to measure how segmented a network actually is.
We define segmentedness as the fraction of potential node-pair communications disallowed by policy (equivalently, the complement of graph edge density) and show it to be the first statistically principled scalar metric for this purpose.
Then, we derive a normalized estimator for segmentedness and evaluate its uncertainty using confidence intervals.
For a 95\% confidence interval with a margin-of-error of $\pm 0.1$, we show that a minimum of $M=97$ sampled node pairs is sufficient.
This result is independent of the total number of nodes in the network, provided that node pairs are sampled uniformly at random.
We evaluate the estimator through Monte Carlo simulations on Erd\H{o}s--R\'enyi, stochastic block models, and real-world enterprise network datasets, demonstrating accurate estimation.
Finally, we discuss applications of the estimator, such as baseline tracking, zero trust assessment, and merger integration.
\end{abstract}


\section {Introduction} \label{sec:intro}

Network segmentation is a widely recommended security practice for limiting lateral movement and reducing the impact of breaches.
Modern security frameworks routinely advise practitioners to divide networks into smaller, isolated zones and tightly control communication between them.
Despite this emphasis, security teams lack a simple way to assess how segmented a network is.
In practice, segmentation is usually assessed qualitatively using architecture diagrams, network security policy reviews, or expert judgment.
This gap exemplifies a broader challenge identified by the 2025 National Academies Cyber Hard Problems report: the absence of security metrics that are reliable, quantifiable, and repeatable \cite{national2025cyber}.

The lack of a quantitative measure has concrete consequences.
Organizations cannot objectively compare segmentation across business units, validate that architecture changes improved isolation, or determine whether network complexity is justified by security benefits.
To our knowledge, no prior work has provided a single, interpretable scalar metric for segmentedness: leaving practitioners unable to objectively measure a practice that security frameworks universally recommend.
Further, practitioners are left without objective evidence to support segmentation decisions or investments.

In this paper, we model a network as a graph and study segmentedness as a measurable property of that graph.
We show that a simple statistic derived from the network's global edge density captures meaningful information about network flatness.
We provide a principled method to estimate this quantity from observed data and attach confidence intervals to the estimate.
Our approach is intentionally lightweight and does not require detailed knowledge of application semantics or traffic intent.
Through analysis and simulation, we demonstrate that the estimator is accurate and well-behaved (unbiased with correct coverage) across a range of network models.
This provides security practitioners with a practical, interpretable metric for day-to-day use.

The rest of the paper is organized as follows.
Section~\ref{sec:related} reviews prior work on network segmentation and graph-based metrics.
Section~\ref{sec:model} introduces our network model and formally defines segmentedness in terms of global edge density.
Section~\ref{sec:estimation} presents our estimator and derives confidence intervals for segmentedness from network data.
Section~\ref{sec:size} provides example sample size calculations for practical deployment.
Section~\ref{sec:evaluation} evaluates the estimator using synthetic networks as well as enterprise networking data.
Section~\ref{sec:bayesian} extends the estimation framework to handle highly segmented networks where the confidence interval degenerates.
Section~\ref{sec:operation} discusses some operational use cases using the segmentedness metric.
Section~\ref{sec:sampling} discusses practical considerations for sampling in operational environments.
Section~\ref{sec:limitation} makes explicit the limitations of our approach.
Section~\ref{sec:conclusion} concludes with a summary of our findings.

\section {Related Work} \label{sec:related}

We review prior work on network segmentation and graph-based measures potentially related to segmentedness.
None of the works reviewed below set out to create a segmentedness (or flatness) measure for security practitioners.
As such, our review should be seen as a summary of prior research focus.

\subsection {Government and Industry Frameworks}

NIST SP~800-207 \cite{nist800207} defines network segmentation as a core design principle of zero trust architectures, emphasizing that access between resources should be tightly controlled and continuously verified.
However, the framework describes segmentation in architectural and qualitative terms, such as the presence of trust boundaries and enforcement points.
As a result, segmentation assessment relies on expert judgment, not on a quantitative measure of the degree of segmentedness or network flatness.

Similarly, CISA Zero Trust Maturity Model \cite{cisa2023ztmm} defines maturity across a four-stage ordinal scale from ``traditional'' to ``optimal'' implementations.
While this model provides a strategic roadmap, it relies on qualitative self-assessment and high-level checklists to determine a network's maturity level.
Consequently, an organization's placement on this scale remains a subjective determination rather than a reflection of a verifiable metric.

NIST CSF~2.0 \cite{nistcsf20} treats network segmentation as an important security control within its broader guidance for managing cybersecurity risk.
The framework evaluates segmentation through qualitative categories and maturity-based profiles, where organizations assess whether network security policies are in place.
CSF~2.0 also relies on expert self-assessment and descriptive tiers, making it difficult to quantitatively compare segmentation strength over time or across organizations.

Taken together, these government and industry frameworks emphasize the importance of segmentation but assess it through qualitative judgment, highlighting the absence of a quantitative, repeatable scalar measure for evaluating how segmented a network is or how that segmentedness changes over time.

\subsection {Network Segmentation Research}

Because network segmentation spans multiple disciplines and relatively few researchers study it directly, we review the available literature comprehensively rather than selectively.

\cite{wagner2016towards} evaluates network segmentation by measuring how different segmentation designs affect both security and the ability of a network to carry out its intended operational tasks.
The paper introduces a unified quantitative metric that combines a security score, representing how often devices remain usable during an attack, with a delay score, representing how much normal network operations are slowed by the attack.
This metric, therefore, measures the practical effectiveness of segmentation in maintaining secure and timely network operation.
It does not directly measure how segmented or flat the network is.

\cite{sabur2019s3} introduces a quantitative Segmentation Index (SI) that is used to evaluate and select a good way of dividing a network into segments.
The index measures how well machines within the same segment belong together and how clearly different segments are separated by firewall rules.
However, SI does not (and was not designed to) indicate how flat or segmented the network is overall.

\cite{osman2020transparent} and \cite{basta2022towards} both evaluate network micro-segmentation through its security effects rather than by defining a single direct measure of segmentedness.
\cite{osman2020transparent} compares different segmentation setups in smart home IoT networks and shows that stronger segmentation leads to reduced attack surface, lower vulnerability exposure, and fewer compromised devices.
Similarly, \cite{basta2022towards} assesses segmentation by measuring how it reduces network exposure and attacker capabilities using multiple graph-based metrics such as connectivity, reachability, centrality, and attack-path characteristics.
In both works, segmentedness is inferred from improved security.
Segmentedness is not captured by a single, scalar metric.

\cite{mhaskar2021formal} gives a precise mathematical definition of network segmentation based on how similar or different machines' access-control rules are, enabling automated construction and validation of segmentation designs.
Under this framework, a network is treated as either correctly segmented or not.
While this binary view is useful for formal reasoning, it does not reflect real networks, where segmentation is often uneven and incremental rather than all-or-nothing.

\cite{tyagi2025measuring} introduces LMS$_k$ as a family of metrics parameterized by the maximum number of attacker steps $k$, enabling analysis of how attacker effectiveness increases as deeper lateral movement is permitted.
For a fixed $k$, LMS$_k$ is computed by estimating, for every pair of distinct machines, the likelihood that an attacker who compromises one can reach the other by chaining together up to $k$ plausible movement steps, with each step weighted by the ease with which the underlying connections can be abused, and then averaging these likelihoods across all pairs.
Lower values of LMS$_k$ correspond to networks in which compromise is difficult to propagate under the assumed attacker model, while higher values indicate environments in which lateral movement is broadly feasible once an initial foothold is obtained.
Although LMS$_k$ was not proposed as a direct measure of segmentation, it provides useful insight into the security consequences of network structure under varying attacker capabilities.
At the same time, the dependence on an explicit attacker model, the need in practice to choose a fixed $k$ (the result converges as $k$ increases, but the computation is cubic in the number of nodes in the network per step), and the absence of a single scalar summary limit its suitability as a standalone measure of segmentedness.

In \cite{bredesen2025network}, the authors experimentally demonstrate the security benefits of segmentation by comparing a flat network architecture against a segmented design during a simulated attack.
Their results quantify the reduction in successful lateral movement when firewalls are introduced between zones.
However, the methodology treats segmentation as a binary condition.
Thus, the approach validates the utility of segmentation, but does not provide a mechanism to measure the relative degree of isolation in more complex, partially segmented environments.

\cite{mujib2020performance} evaluates micro-segmentation by measuring its impact on network performance, showing that enabling segmentation adds only minimal delay and no packet loss.
\cite{kotha2020network} discusses network segmentation as a security best practice and evaluates it qualitatively through design principles and case studies.
\cite{yatagha2023security} focuses on automatically creating segments in IIoT networks using traffic clustering and measures the quality of the clustering.
\cite{alofeishat2024build} compares different segmentation and micro-segmentation deployment approaches using high-level operational factors such as security level, cost, and complexity.
\cite{mani2025securing} evaluates role-based micro-segmentation in public clouds using metrics like role inference accuracy and policy violations, focusing on whether segmentation rules are applied correctly rather than on how segmented the network is.
\cite{dube2026taxonomy} provides a comprehensive taxonomy of segmentation concepts and practices across industry and research, documenting the diverse ways in which segmentation is defined and measured, and observes the lack of a generally accepted scalar metric for measuring the degree of segmentation or network flatness.

In summary, while recent literature has transitioned from qualitative best practices to quantitative evaluation, a significant gap remains.
The most advanced current approaches, such as those in \cite{basta2022towards} and \cite{tyagi2025measuring}, provide measures of risk such as exposure, robustness, and lateral-movement susceptibility.
However, these metrics focus on attack outcomes, estimating how likely specific attacks are to succeed in a given network.
To our knowledge, no prior work has proposed a scalar metric that quantifies ``segmentedness'' as an intrinsic graph (network) property: independent of specific threat models or assumptions about how likely individual services are to be compromised.
Such a metric would allow practitioners to answer the fundamental question, ``How segmented is my network?'' by providing a direct measure of network segmentedness (flatness) that can be tracked consistently across different designs and over time.

\subsection {Graph-based Measures} \label{sec:graph-based}

While much of the prior work on network segmentation treats segmentation as a binary architectural or policy correctness property, graph-based measures provide quantitative characterizations of network structure that are related to segmentation as understood in security practice.
These measures capture properties such as structural cohesion and separability, which are relevant to understanding certain aspects of network organization and risk.
They have been applied to graph analysis and represent natural candidates for quantifying network segmentation.
However, as discussed below, thus far they have not been adapted to directly quantify the overall permissiveness of communication in a network.

\emph{Edge density} is a simple graph-based measure \cite{wasserman1994social} that quantifies how many connections exist in a graph relative to how many connections are possible.
Given a graph with $n$ nodes, there are $\binom{n}{2}$ possible pairwise connections, and the edge density is defined as the fraction of those possible connections that are present.
An edge density of one corresponds to a fully connected graph in which every node can communicate with every other node, while an edge density of zero corresponds to a graph with no allowed communication between nodes.
Because it directly reflects the proportion of permitted communication relationships, edge density provides an intuitive measure of how flat (interconnected) a network is, independent of how those connections are arranged.
Although prior work has not proposed edge density as a segmentedness metric, it provides a natural foundation for such a measure.

\emph{Modularity} is another graph-based measure that evaluates how well a graph can be divided into groups, or communities, whose nodes are more densely connected to each other than to the rest of the graph~\cite{newman2004finding}.
Intuitively, modularity compares the number of connections within proposed groups to the number that would be expected if connections were placed at random, while preserving overall connectivity levels.
High modularity values indicate that the graph has a clear community structure, with relatively strong separation between groups, whereas low modularity values suggest that connections are more evenly distributed and that the graph lacks meaningful subdivision.
As such, modularity is useful for identifying and evaluating segmentation in terms of clustered structure, but it emphasizes the arrangement of connections.
In contrast, edge density directly reflects how many communication relationships are permitted overall, independent of how those relationships are grouped.

The \emph{Fiedler value}, also known as \emph{algebraic connectivity}, is a numerical property of a graph that captures how strongly connected the graph is as a whole and how difficult it is to separate into disconnected components~\cite{fiedler1973algebraic}.
A value of zero indicates that the graph is disconnected, while larger values indicate increasing resistance to separation, with graphs containing bottlenecks or hub-and-spoke structures typically exhibiting low Fiedler values.
As such, the Fiedler value is relevant to network segmentedness when segmentedness is interpreted in terms of structural separability or vulnerability to cuts.
However, it is less suitable when segmentedness is interpreted as the overall permissiveness of communication, since graphs with the same number of allowed connections can have very different Fiedler values depending on how those connections are arranged.

Despite their relevance, none of these graph-based measures have been adapted to quantify network segmentedness in a statistically principled, network-size-independent manner.
Our work is the first to bridge this gap.

\section {Formal Model} \label{sec:model}

We propose defining segmentedness as the overall permissiveness of communication policy, formalized through edge density.
Accordingly, we model segmentedness using edge density, which directly captures the fraction of end-to-end communication relationships that are permitted and serves as the foundation for the formal model developed in the remainder of this section.

We acknowledge that the structural arrangement of permitted connections can influence security outcomes, for example, by creating chokepoints or enabling indirect communication paths.
However, as documented in Section~\ref{sec:related}, security practitioners currently lack a simple quantitative metric that summarizes how permissive a network's communication policy is; our objective prioritizes a single, easily interpretable scalar over metrics that emphasize structure.
This objective explicitly rules out structure-oriented measures such as modularity and the Fiedler value, which capture how connections are arranged rather than how much communication is permitted overall (see Appendix~\ref{app:comparison} for a comparison of edge density, modularity and the Fiedler value).

\subsection {Network Model} \label{sec:network-model}

We model a network as an undirected simple graph $G = (V, E)$, where $V$ is the set of nodes representing end systems such as user devices or server workloads, and $E \subseteq \{\{u,v\} \mid u,v \in V, u \neq v\}$ is the set of edges.
An edge $\{u,v\} \in E$ indicates that end-to-end communication between nodes $u$ and $v$ is permitted by network security policy, independent of the physical or logical path taken through the underlying network.
A missing edge between two nodes means that direct communication between them is not permitted by policy.

Network infrastructure components, including routers, switches, firewalls, and other middle-boxes, are not represented explicitly as nodes.
Edges represent policy-permitted end-to-end communication between end systems, abstracting away the physical or logical path taken through the network infrastructure.

Edges are assumed to be bidirectional, reflecting symmetric communication capability.
As $G$ is a simple graph, self-loops and parallel edges are not allowed. \fnmark{bidirectional}

Let $n = |V|$ denote the total number of nodes in the network.
The maximum number of possible edges is $\binom{n}{2}$, corresponding to a fully connected network in which all pairs of nodes are allowed to communicate.

\fntext {bidirectional}{
    We focus on segmentation within an organization's internal network rather than perimeter security.
    While firewall rules can be asymmetric, our undirected edge model is appropriate for internal segmentation, as internal network security (access control) policies are often symmetric.
    In this paper, policy symmetry is a modeling assumption.
}

\subsection {Flatness and Segmentedness}

We define \emph{flatness} as a normalized measure of how permissive the network communication policy is.
Specifically, the flatness of a network graph $G$ is defined as
\begin{equation}
F(G) = \frac{|E|}{\binom{n}{2}},
\end{equation}
that is, the fraction of all possible node pairs for which communication is allowed.

We define \emph{segmentedness} as the complement of flatness:
\begin{equation}
S(G) = 1 - F(G).
\end{equation}
Segmentedness, therefore, quantifies the proportion of potential communication relationships that are disallowed by policy.

We note that the term ``segmentation'' is used in various ways across industry and research, with no single universally accepted definition \cite{dube2026taxonomy}.
In this paper, we use the term \emph{segmentedness} to refer specifically to the metric defined above: the proportion of potential node-pair communications that are denied by policy.
This operational definition allows us to provide a quantitative, comparable measure without requiring consensus on the broader conceptual boundaries of what constitutes ``segmentation.''

\subsection {Metric Scope}

The segmentedness metric measures how permissive network communication policies are between pairs of nodes.
It captures exposure at the policy level by considering whether direct communication between two nodes is allowed.

The metric does not describe network architecture, performance, or implementation details.
It does not model routing or switching topology, traffic engineering, or the correctness of policy enforcement.
It does not account for transitive compromise, multi-hop lateral movement, or attacker behavior beyond the existence of permitted direct communication.

\subsection {Properties} \label{sec:properties}
The segmentedness metric satisfies several properties useful for comparing network security policies:
\begin{itemize}
\item \textbf{Boundary conditions:} A fully connected network has segmentedness $S(G)=0$, while a fully disconnected network has segmentedness $S(G)=1$.
\item \textbf{Monotonicity:} Adding an allowed communication edge to the network strictly decreases segmentedness, while removing an allowed edge strictly increases segmentedness.
\item \textbf{Scale invariance:} Segmentedness is normalized and can be compared across networks of different sizes.
\item \textbf{Structural neutrality:} Segmentedness depends only on the number of permitted communication relationships, not on their arrangement within the network.
\item \textbf{Interpretability:} Segmentedness has a probabilistic interpretation: $S(G)$ represents the probability that two uniformly random nodes cannot communicate directly.
The probabilistic interpretation holds when a node-pair is selected uniformly at random from the set of all $\binom{n}{2}$ unordered pairs.
\end{itemize}

\subsection {Expected Number of Reachable Neighbors} \label{sec:reachable-pairs}

The expected number of directly reachable neighbors of a randomly chosen node depends solely on the edge density and the number of nodes in the network.

Select a node $v \in V$ uniformly at random, and let $N(v)$ denote the set of nodes that are directly reachable from $v$, i.e., the neighbors of $v$.
The expected size of $N(v)$ is the average node degree in $G$:
\begin{equation}
\mathbb{E}[|N(v)|] = \frac{1}{n} \sum_{w \in V} \deg(w).
\end{equation}
In any undirected graph, each edge contributes exactly one to the degree of each of its two endpoints, so the sum of all node degrees equals twice the number of edges: $\sum_{w \in V} \deg(w) = 2|E|$.
Substituting this identity and $|E| = \binom{n}{2} F(G) = \frac{n(n-1)}{2} F(G)$ yields
\begin{equation}
\mathbb{E}[|N(v)|] = \frac{2|E|}{n} = (n-1)\, F(G).
\end{equation}

Since edges represent policy-permitted end-to-end communication, $\mathbb{E}[|N(v)|] = (n-1)F(G)$ can be interpreted as the expected \emph{direct exposure} of a randomly chosen node: on average, a node is permitted to communicate with $(n-1)F(G)$ other nodes.
Consequently, for fixed network size $n$, reducing edge density proportionally reduces the expected number of systems that are directly reachable from any given system, which in turn reduces the expected set of immediate communication partners available to an attacker who obtains an initial foothold on a random node.

\section{Empirical Estimation Methodology} \label{sec:estimation}

We now present a practical methodology for estimating segmentedness in large networks when complete knowledge of network connectivity is unavailable.
Rather than attempting to enumerate all permitted communication relationships, the approach relies on randomized sampling and empirical connectivity testing to estimate segmentedness with statistical confidence.

\subsection {Measurement Challenge}
Direct computation of segmentedness requires knowledge of all permitted communication pairs in the network, which is infeasible in large enterprise environments.
For a network with $n$ nodes, there are $\binom{n}{2}$ possible node pairs, making exhaustive testing impractical even for moderately sized networks.
In addition, network security policies are often distributed across systems, further complicating direct analysis.

As a result, segmentedness must be estimated empirically using partial observations, while providing explicit bounds on estimation error.

\subsection {Randomized Sampling of Node Pairs}
To estimate $F(G)$, we randomly sample $M$ unordered pairs of distinct nodes $\{u,v\}$ from $V$ with replacement.
For each sampled pair, we perform a connectivity test to determine whether communication between $u$ and $v$ is permitted by policy.
Each test yields a Bernoulli outcome:
\begin{equation}
X_i =
\begin{cases}
1 & \text{if communication is permitted}, \\
0 & \text{otherwise}.
\end{cases}
\end{equation}

The empirical flatness estimate is then given by
\begin{equation}
\hat{F} = \frac{1}{M} \sum_{i=1}^{M} X_i,
\end{equation}
with corresponding segmentedness estimate $\hat{S} = 1 - \hat{F}$.

\subsection {Connectivity Test Suite}
Determining whether communication between two nodes is permitted requires operational testing.
Relying on a single test, such as ICMP echo, can produce false negatives due to protocol-specific filtering or configuration.
To improve robustness, we propose a battery of connectivity tests that may include ICMP probes, UDP packet exchanges, and TCP connection attempts.

A node pair is considered connected if \emph{any} test in the suite succeeds.
This definition reflects the security-relevant observation that the existence of at least one usable communication channel is sufficient for interaction between systems.
The test suite is fixed for all measurements to ensure consistency and comparability across sampling runs.

\subsection {Statistical Guarantees and Sample Size} \label{sec:guarantee}
Each sampled pair test can be viewed as an independent Bernoulli trial with success probability $p = F(G)$.
Under this model, the estimator $\hat{F}$ is an unbiased estimator of $p$ with variance
\begin{equation}
\mathrm{Var}(\hat{F}) = \frac{p(1-p)}{M}.
\end{equation}

By the central limit theorem \cite{feller1991introduction}, for sufficiently large $M$ (typically $M \geq 30$ suffices), $\hat{F}$ is approximately normally distributed.
An approximate two-sided confidence interval with confidence level $1-\alpha$ is given by
\begin{equation}
\hat{F} \pm z_{\alpha/2} \sqrt{\frac{p(1-p)}{M}},
\end{equation}
where $z_{\alpha/2}$ is the corresponding quantile of the standard normal distribution.
This is known as the Wald or normal approximation confidence interval~\cite{agresti1998approximate}.

Since $p$ is unknown, a conservative bound can be obtained by noting that $p(1-p)$ is maximized at $p = 0.5$.
This yields a worst-case standard error of
\begin{equation}
\mathrm{SE}(\hat{F}) \le \frac{1}{2\sqrt{M}}.
\end{equation}
This worst-case assumption means that the required sample size $M$ is valid for any network, regardless of its actual segmentedness level.

To achieve an absolute error of at most $\varepsilon$ with confidence level $1-\alpha$, it suffices to choose
\begin{equation}
M \ge \frac{z_{\alpha/2}^2}{4\varepsilon^2}.
\label{eq:sample_size}
\end{equation}

Because segmentedness is defined as $S(G) = 1 - F(G)$, the same confidence interval and error bounds apply directly to $\hat{S}$.
Importantly, the required sample size depends only on the desired accuracy and confidence level, and not on the total number of nodes in the network, provided that node pairs are sampled uniformly at random. \fnmark{replacement}
Compared to edge density, modularity and the Fiedler value (Section~\ref{sec:graph-based}) are significantly more challenging to estimate via sampling.

\fntext {replacement} {
  Independence holds under sampling with replacement (as stated earlier in the paper).
  When sampling without replacement, a finite-population correction factor of $\sqrt{1 - M/\binom{n}{2}}$ applies to the standard error.
  For all network sizes considered in this paper, this correction is negligible: even for the smallest graph evaluated ($n = 86$, $\binom{n}{2} = 3{,}655$), the correction factor exceeds $0.98$.
}

\section{Example Sample Size Calculations} \label{sec:size}

Using the formula derived in Section~\ref{sec:guarantee}, we compute the minimum number of randomly sampled node pairs $M$ required to estimate segmentedness with a specified confidence interval.
We consider a two-sided 95\% confidence interval (CI) with half-width $\varepsilon = 0.1$.

For a 95\% confidence level, the corresponding normal quantile is $z_{0.025} \approx 1.96$.
Applying the conservative bound that maximizes variance at $p=0.5$, the required sample size satisfies
\begin{equation}
M \ge \frac{z_{0.025}^2}{4\varepsilon^2}.
\end{equation}
Substituting the desired values yields
\begin{equation}
M \ge \frac{(1.96)^2}{4(0.1)^2}
= \frac{3.8416}{0.04}
= 96.04.
\end{equation}
Rounding up to the nearest integer, this gives a minimum of $M = 97$ sampled node pairs.

Table~\ref{tab:sample-size-01} summarizes the corresponding conservative estimates of $M$ for networks with 1{,}000, 10{,}000 and 100{,}000 nodes under the same confidence requirements.
Tables~\ref{tab:sample-size-01-multi} and \ref{tab:sample-size-005-multi} show the estimates for $M$ for networks with 1{,}000, 10{,}000, and 100{,}000 nodes, when the confidence coefficient and the half-width are varied.

That a 95\% CI with a half-width of $\pm 0.1$ can be achieved with just $97$ node pairs is an elegant result: it makes the use of edge density as a measure of segmentedness feasible for networks of arbitrary size.
Segmentedness measurements can be repeated frequently (for example, every quarter), even if they have to be run manually or over low-bandwidth links.
See Section~\ref{sec:bayesian} for handling of the special case when the empirical number of connections is zero.

\begin{table}[htbp]
\centering
\caption{Conservative minimum number of randomly sampled node pairs $M$ required to estimate segmentedness with a 95\% CI of $\pm 0.1$, shown alongside the maximum number of possible edges for each network size.}
\label{tab:sample-size-01}
\begin{tabular}{r r r}
\hline
\textbf{\# nodes} & \textbf{Max. possible edges} & \textbf{Min. $M$ 95\% CI $\pm 0.1$} \\
\hline
$1{,}000$   & $499{,}500$              & $97$ \\
$10{,}000$  & $49{,}995{,}000$         & $97$ \\
$100{,}000$ & $4{,}999{,}950{,}000$    & $97$ \\
\hline
\end{tabular}
\end{table}

\begin{table}[htbp]
\centering
\caption{Conservative minimum number of randomly sampled node pairs $M$ required to estimate segmentedness with a CI of $\pm 0.1$ at 90\%, 95\%, and 99\% confidence levels.}
\label{tab:sample-size-01-multi}
\begin{tabular}{r r r r}
\hline
\textbf{\# nodes} & \textbf{$M$ (90\% CI)} & \textbf{$M$ (95\% CI)} & \textbf{$M$ (99\% CI)} \\
\hline
$1{,}000$   & $68$  & $97$  & $166$ \\
$10{,}000$  & $68$  & $97$  & $166$ \\
$100{,}000$ & $68$  & $97$  & $166$ \\
\hline
\end{tabular}
\end{table}

\begin{table}[htbp]
\centering
\caption{Conservative minimum number of randomly sampled node pairs $M$ required to estimate segmentedness with a CI of $\pm 0.05$ at 90\%, 95\%, and 99\% confidence levels.}
\label{tab:sample-size-005-multi}
\begin{tabular}{r r r r}
\hline
\textbf{\# nodes} & \textbf{$M$ (90\% CI)} & \textbf{$M$ (95\% CI)} & \textbf{$M$ (99\% CI)} \\
\hline
$1{,}000$   & $271$ & $385$ & $664$ \\
$10{,}000$  & $271$ & $385$ & $664$ \\
$100{,}000$ & $271$ & $385$ & $664$ \\
\hline
\end{tabular}
\end{table}

\section {Estimator Evaluation} \label{sec:evaluation}

We evaluate the proposed estimation procedure through Monte Carlo simulations on synthetic networks and via a real-world network dataset.
We begin with the Erd\H{o}s--R\'enyi model to establish baseline estimator behavior, then assess robustness under the stochastic block model, which introduces heterogeneous connectivity patterns representative of segmented enterprise networks.
Finally, we apply the estimator to real-world communication graphs from an enterprise networking dataset to validate its performance on observed enterprise traffic.

\subsection {Estimation over Random Networks} \label{sec:er}

We first evaluate the proposed estimation procedure in a simple and well-understood random network setting.
The Erd\H{o}s--R\'enyi model provides a natural baseline, as it captures networks in which connections between nodes occur independently and uniformly at random, while still allowing the overall level of connectivity to be controlled by a single parameter.
This makes it a useful reference point for understanding estimator behavior before considering more structured graph models.

In the Erd\H{o}s--R\'enyi model $G(n,p)$ \cite{bollobas2011random}, a graph is generated on $n$ labeled nodes by including each unordered pair of distinct nodes as an edge independently with probability $p$.
In this setting, the edge probability $p$ coincides with the true global edge density $F(G)$.
Rather than explicitly constructing the full graph, the estimation procedure samples $M$ unordered node pairs uniformly at random and observes whether an edge is present.
Each observation corresponds to a Bernoulli random variable $X_i \sim \mathrm{Bernoulli}(F(G))$, and the estimator $\hat{F} = \frac{1}{M}\sum_{i=1}^M X_i$ coincides with the empirical edge density defined earlier.

\begin{figure} [htbp]
  \centering
  \includegraphics[width=0.5\textwidth]{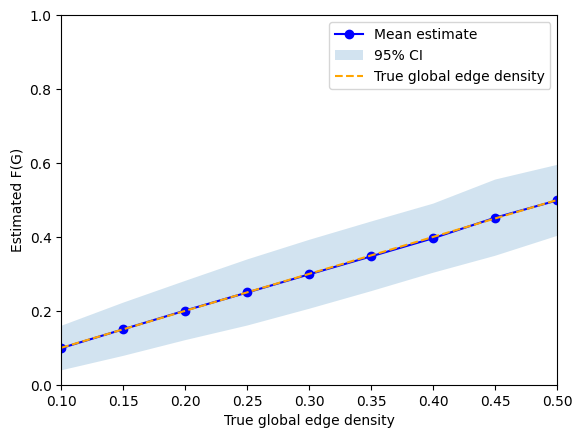}
  \caption{Monte Carlo mean and 95\% CI of the edge density estimator under the Erd\H{o}s--R\'enyi $G(n,p)$ model as a function of the true global edge density; $M=97$.}
  \label{fig:er-estimate}
\end{figure}

Figure~\ref{fig:er-estimate} illustrates the Monte Carlo behavior of the estimator $\hat{F}$ for Erd\H{o}s--R\'enyi graphs with $n=100{,}000$ nodes, chosen to mimic the scale of a large organizational network.
The true edge probability $p$ is varied over the range $0.1 \le p \le 0.5$, corresponding to different levels of network connectivity.
For each value of $p$, the markers show the mean estimate of $\hat{F}$ across repeated ($1{,}000$) trials, and the shaded region indicates the spread of estimates across Monte Carlo simulations.
The estimates lie close to the diagonal, where estimate equals true value, across all values of $p$, confirming that the estimator is unbiased in this setting and behaves stably.

\begin{figure} [htbp]
  \centering
  \includegraphics[width=0.5\textwidth]{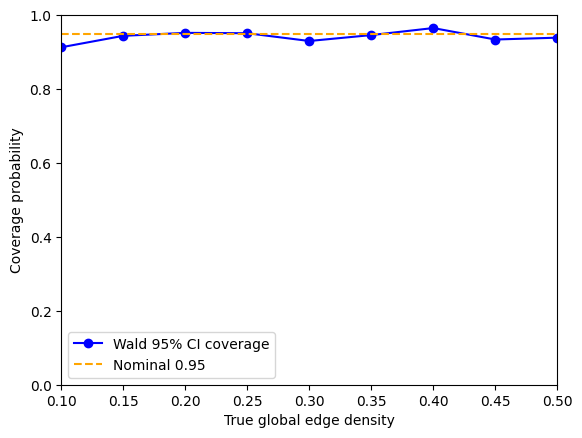}
  \caption{Empirical coverage probability of the 95\% Wald confidence interval for the edge density estimator under the Erd\H{o}s--R\'enyi $G(n,p)$ model as a function of the true global edge density; $M=97$}
  \label{fig:er-coverage}
\end{figure}

Figure~\ref{fig:er-coverage} reports the empirical coverage probability of the $95\%$ Wald confidence interval for the same Erd\H{o}s--R\'enyi simulations.
For each value of $p$, coverage is computed as the fraction of Monte Carlo trials in which the confidence interval contains the true edge density.
Across the full range of connectivity levels considered, the observed coverage remains close to the nominal level of $0.95$, indicating that the confidence intervals used in this paper are well-behaved under the Erd\H{o}s--R\'enyi baseline.

\subsection {Estimation over Networks with Communities} \label{sec:sbm-equal}

We next evaluate the proposed estimation procedure in a random network model with explicit community structure.
The stochastic block model (SBM) provides a natural extension of the Erd\H{o}s--R\'enyi baseline by allowing edge probabilities to vary across different classes of node pairs \cite{holland1983stochastic}.
This setting makes it possible to assess the robustness of the estimator when global edge density is determined by local connectivity patterns.

In the SBM with $K$ equal-sized blocks, a graph is generated on $n$ labeled nodes by first partitioning the nodes into $K$ disjoint groups of equal size.
Edges between nodes in the same block are included independently with probability $p_{\text{in}}$, while edges between nodes in different blocks are included independently with probability $p_{\text{out}}$.
Although all edge decisions remain independent, the probability of an edge now depends on the block memberships of the node pair.
The resulting global edge density is a weighted average of $p_{\text{in}}$ and $p_{\text{out}}$ for large $n$:
\begin{equation}
F(G) = \frac{1}{K} p_{\text{in}} + \frac{K-1}{K} p_{\text{out}},
\end{equation}
where the weights $\frac{1}{K}$ and $\frac{K-1}{K}$ represent the proportions of within-block and between-block node pairs, respectively. \fnmark{exact}

The same sampling procedure from Section~\ref{sec:er} is applied here, with each observation being a Bernoulli trial with success probability equal to the global edge density of the SBM.

\begin{figure} [htbp]
  \centering
  \includegraphics[width=0.5\textwidth]{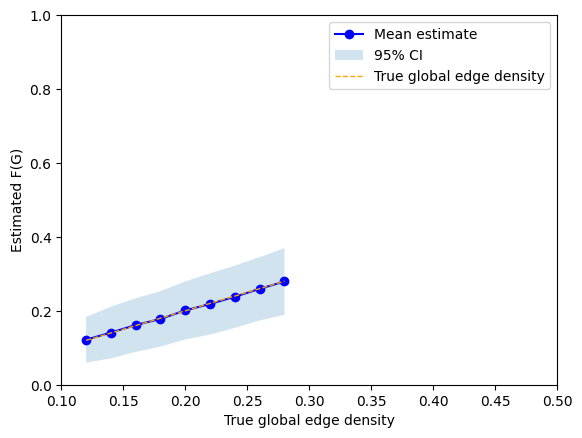}
  \caption{Monte Carlo mean and 95\% CI of the edge density estimator under the SBM equal-sized blocks model (p\_out=0.1) as a function of the true global edge density; $M=97$.}
  \label{fig:sbm-equal-0.1}
\end{figure}

\begin{figure} [htbp]
  \centering
  \includegraphics[width=0.5\textwidth]{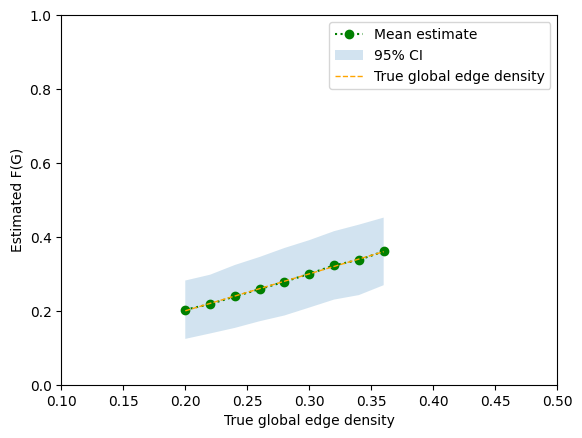}
  \caption{Monte Carlo mean and 95\% CI of the edge density estimator under the SBM equal-sized blocks model (p\_out=0.2) as a function of the true global edge density; $M=97$.}
  \label{fig:sbm-equal-0.2}
\end{figure}

Figures~\ref{fig:sbm-equal-0.1} and~\ref{fig:sbm-equal-0.2} illustrate the Monte Carlo behavior of the estimator $\hat{F}$ for SBM networks with $n=100{,}000$ nodes and $K=5$ equal-sized blocks.
In each figure, the between-block probability $p_{\mathrm{out}}$ is held fixed, with $p_{\mathrm{out}}=0.1$ in Figure~\ref{fig:sbm-equal-0.1} and $p_{\mathrm{out}}=0.2$ in Figure~\ref{fig:sbm-equal-0.2}, while the within-block probability $p_{\mathrm{in}}$ is varied to produce global edge densities in the range $[0.1, 0.5]$, similar to Section~\ref{sec:er}.
For each parameter configuration, the markers show the mean of $\hat{F}$ across $1{,}000$ Monte Carlo trials, and the shaded region represents the associated $95\%$ confidence interval.
Across the full range of edge densities considered, the estimates closely follow the diagonal, on either side, indicating that the estimator remains unbiased in the presence of community structure.

\begin{figure} [htbp]
  \centering
  \includegraphics[width=0.5\textwidth]{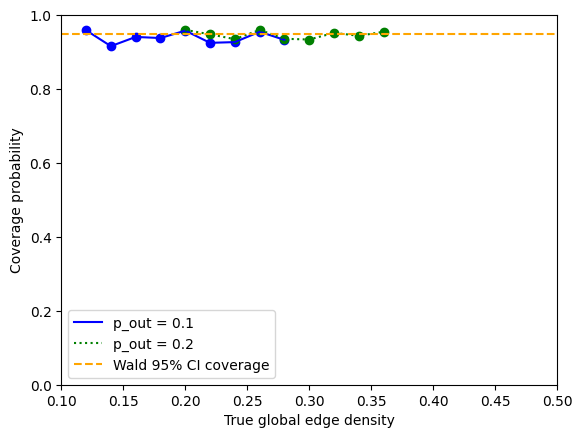}
  \caption{Empirical coverage probability of the 95\% Wald confidence interval for the edge density estimator under the SBM equal-sized model as a function of the true global edge density; $M=97$.}
  \label{fig:sbm-equal-coverage}
\end{figure}

Figure~\ref{fig:sbm-equal-coverage} reports the empirical coverage probability of the $95\%$ Wald confidence interval for the same SBM simulations.
Coverage is computed as the fraction of Monte Carlo trials in which the confidence interval contains the true global edge density.
For both values of $p_{\mathrm{out}}$, the observed coverage remains close to the nominal level of $0.95$ across the range of edge densities examined.
These results suggest that the Wald confidence intervals used in this paper remain well-behaved for edge density estimation under equal-sized stochastic block models.

The SBM results demonstrate that the sampling-based estimation procedure remains accurate even when network structure exhibits significant heterogeneity, as would be expected in real enterprise networks with business-unit or functional segmentation.

\fntext {exact} {
  The exact expression for $K$ equal-sized blocks of size $n/K$ is $F(G) = \frac{n/K - 1}{n - 1}\,p_{\text{in}} + \frac{n - n/K}{n - 1}\,p_{\text{out}}$, which converges to the stated form as $n \to \infty$.
  For $n = 100{,}000$ and $K = 5$, the approximation error is negligible.
}

\subsection{Estimation over Enterprise Network Data} \label{sec:cisco}

While the previous simulations demonstrate the estimator's performance on synthetic structures like the SBM, we now evaluate its utility on actual enterprise network topologies.

The publicly available Cisco Secure Workload enterprise networking dataset \cite{madani2022dataset, snap_cisco_data} provides a unique opportunity to evaluate the estimator on realistic enterprise connectivity patterns.
These graphs are constructed from observed network traffic recorded by sensors over several days, rather than from static security policies or firewall access control lists.
Consequently, the edges represent active communications that actually occurred during the observation window.
While these graphs provide a realistic snapshot of network activity, they do not necessarily capture the full reachability matrix that a network security policy might permit.

In a well-segmented environment, graphs representing policy-permitted communication will generally be denser than those built from observed traffic over a few days.
This is because many permitted pathways, such as disaster recovery routes or rarely used management ports, may remain idle during a short observation period.
Indeed, the majority of the graphs in the enterprise networking dataset exhibit extreme sparsity, with global edge densities $F(G)$ typically falling well below 0.05 (in calculating true edge density of these networks, we count any communication between two nodes as evidence of a link between the two nodes).

Although the enterprise networking dataset reflects observed traffic rather than configured policy, the statistical properties of the estimator---unbiasedness and coverage---depend only on the sampling procedure and the true edge density of the underlying graph, regardless of whether that graph represents policy or observed communication.

For this evaluation, we selected the five densest graphs in the dataset to approximate real-world networks and network security policies.
These are the $g1$, $g11$, $g19$, $g21$, and $g7$ enterprise networks.
As shown in Table \ref{tab:cisco-results}, the empirical mean edge density $\hat{F}$ calculated over 1,000 Monte Carlo trials matches the true edge density $F(G)$ closely.
The true density remains within the calculated 95\% confidence intervals for all graphs, empirically validating the estimator's accuracy and its independence from the total node count $n$.

These results demonstrate that the estimator performs accurately on real enterprise topologies with diverse connectivity patterns, validating its practical applicability beyond synthetic models.

\begin{table}[htbp]
\centering
\caption{Network Metrics and Estimator Performance for the five densest graphs from the enterprise networking dataset ($M=97$, $1,000$ trials.)}
\label{tab:cisco-results}
\footnotesize
\begin{tabular*}{\columnwidth}{@{\extracolsep{\fill}}lrrrrr@{}}
\toprule
\textbf{Graph} & \textbf{$n$} & \textbf{$|E|$} & \textbf{$F(G)$} & \textbf{Mean $\hat{F}(G)$} & \textbf{95\% CI} \\
\midrule
g1         &  1447 &  106617 &    0.102 &      0.102 & [0.101, 0.104] \\
g11        &   207 &    1557 &    0.073 &      0.074 & [0.073, 0.076] \\
g19        &    86 &     150 &    0.041 &      0.041 & [0.039, 0.042] \\
g21        &   317 &    1689 &    0.034 &      0.034 & [0.033, 0.035] \\
g7         &   290 &     732 &    0.017 &      0.018 & [0.017, 0.019] \\
\bottomrule
\end{tabular*}
\end{table}

\section {Bayesian Inference for High-segmentedness} \label{sec:bayesian}

The use of the Wald confidence interval in Section~\ref{sec:guarantee} poses a mathematical challenge when the number of observed connections is zero.
In such cases, the empirical flatness $\hat{F}$ becomes zero, which causes the standard error to collapse and results in a zero-width confidence interval.
This zero-width interval is misleading for security practitioners as it suggests absolute certainty that no communication paths exist, when in reality it is merely a result of the empirical sample and the sample size used.

To address this, we introduce a Bayesian Beta-binomial model to provide a more robust estimation of uncertainty at the boundaries~\cite{gelman2013bayesian}.
The Beta distribution is the conjugate prior for the Binomial distribution, making it an ideal choice for updating our beliefs about network flatness based on sampled data.

We incorporate a prior belief that most enterprise networks have an edge density of at least 0.01 (one connection per 100 pairs).
This 1\% baseline reflects the assumption that even highly segmented enterprise networks maintain minimal operational connectivity (e.g., for monitoring, backup, or administrative access).
This prior is modeled using a Beta distribution with hyperparameters $\alpha = 1$ and $\beta = 99$, which centers our initial expectation at 1\%. \fnmark{bayesian}

Following the empirical observation of $k=0$ connections across $M=97$ sampled pairs, we update these hyperparameters to their posterior values.
The posterior distribution becomes $\text{Beta}(\alpha', \beta')$, where the parameters are updated as follows:
\begin{equation}
\alpha' = \alpha + k = 1 + 0 = 1
\end{equation}
\begin{equation}
\beta' = \beta + M = 99 + 97 = 196
\end{equation}
The resulting posterior mean for flatness, representing our new point estimate, is:
\begin{equation}
\hat{F}_{Bayes} = \frac{\alpha'}{\alpha' + \beta'} = \frac{1}{197} \approx 0.005
\end{equation}

Rather than a frequentist confidence interval, we calculate a 95\% credible interval using the quantiles of the posterior Beta distribution.
For this specific case where $\alpha'=1$, the 95\% upper bound for flatness ($F_{upper}$) is given by:
\begin{equation}
F_{upper} = 1 - (0.05)^{1/(\alpha' + \beta' - 1)} = 1 - (0.05)^{1/196} \approx 0.015
\end{equation}
where the upper bound is derived from the 95th percentile of the posterior $\text{Beta}(1, 196)$ distribution.

From a security practitioner's perspective, this means that even with zero observed connections in 97 trials, we can only be 95\% certain that the true network flatness is below 0.015.
This interpretation provides a more realistic and conservative assessment of risk, acknowledging that rare lateral movement paths may still exist despite not being captured in a small sample.
It transforms a mathematically impossible ``perfect'' score into a statistically defensible claim of high segmentedness.

\fntext {bayesian} {
  The posterior is sensitive to the choice of prior: for example, $\text{Beta}(1, 49)$ shifts the posterior mean to approximately $0.01$ and the $95\%$ upper bound to approximately $0.03$.
  We select $\text{Beta}(1, 99)$ as a conservative baseline reflecting minimal operational connectivity, but practitioners should adjust the prior to reflect their environment.
}

\section {Operational Use Cases} \label{sec:operation}

With a well-defined metric in hand, several use cases are possible.

\textbf{Baseline measurement and trend tracking}
A primary operational use case is establishing a baseline segmentedness measurement and tracking changes over time.
For example, an organization with $10{,}000$ endpoints might conduct quarterly measurements using $M=97$ sampled pairs ($95\%$ CI, $\pm0.1$ half-width).
If the initial measurement yields $\hat{S} = 0.3 (\pm0.1)$, indicating $30\%$ of potential communications are blocked, subsequent measurements can track whether segmentation initiatives are improving network isolation.
A decrease to $\hat{S} = 0.2$ in the following quarter would indicate policy drift or new connectivity requirements reducing segmentation levels, prompting investigation.

\textbf{Environment consolidation}
Organizations frequently consolidate previously separate environments, such as development, staging, and production networks, or merge multiple data centers into a shared infrastructure platform.
These initiatives often introduce broad new communication paths between systems that were previously isolated, resulting in large changes to network policy permissiveness.
Segmentedness measurements taken before and after consolidation can quantify the security impact of these architectural decisions and help ensure that consolidation efforts do not unintentionally produce an overly flat network.

\textbf{Merger and acquisition integration}
During mergers or acquisitions, organizations may be required to integrate networks.
The segmentedness metric provides a quantitative proxy for network security during the integration process.
For instance, two networks with pre-merger segmentedness values of $\hat{S_1} = 0.5$ and $\hat{S_2} = 0.4$ might temporarily show $\hat{S}_{combined} = 0.3$ immediately after connection, reflecting increased connectivity between previously isolated systems.
Tracking this metric helps security teams ensure that integration does not inadvertently create an overly flat combined network.

\textbf{Zero Trust implementation assessment}
Organizations implementing zero trust architectures can use segmentedness measurements to validate progress.
A traditional network might measure $\hat{S} = 0.2$, indicating relatively flat connectivity.
As micro-segmentation policies are deployed incrementally as part of a structured zero trust program---first isolating critical assets, then production workloads, then development environments---measurements should show increasing segmentedness (e.g., $ \hat{S} = 0.3, 0.4, 0.5$), providing quantitative evidence of architectural improvement aligned with zero trust principles.
Further, by quantifying the reduction in global edge density, the segmentedness score provides a direct proxy for the difficulty of unauthorized lateral movement.
As the score increases, the number of available reachable paths for an attacker decreases, thereby forcing a reliance on fewer, more heavily monitored chokepoints and significantly raising the effort required to reach critical assets.

\textbf{Benchmarking}
While no established benchmarks currently exist for segmentedness values, the metric enables future development of industry-specific or sector-specific guidelines.
Financial services organizations, which typically enforce stricter isolation requirements, might target $\hat{S} \ge 0.6$, while educational institutions with more open collaboration requirements might operate effectively at $\hat{S} \approx 0.3$.
As organizations adopt this metric, empirical data will enable peer comparison and the establishment of reasonable target ranges.

In each case, the metric's independence from network size and its explicit confidence bounds make repeated measurement practical.
A step-by-step procedure for implementing segmentedness measurements in operational environments is provided in Appendix~\ref{app:deployment}.

\section{Practical Considerations for Sampling} \label{sec:sampling}

The segmentedness metric assumes that node pairs are sampled uniformly at random from the set of networked entities.
In operational environments, implementing such sampling introduces practical challenges that may affect how closely real-world measurements adhere to this assumption.
While Section~\ref{sec:limitation} discusses sampling limitations from a statistical perspective, this section focuses on operational challenges that practitioners may encounter when deploying the measurement methodology.

\textbf{Incomplete node enumeration}
Uniform random sampling presumes the ability to enumerate the set of nodes over which communication policies are defined.
In practice, asset inventories may be incomplete or dynamic, particularly in environments with ephemeral virtual machines, containers, or unmanaged endpoints.
As a result, sampling procedures may disproportionately select long-lived or well-managed systems, potentially biasing measurements toward portions of the network with more explicit or carefully maintained policies.

\textbf{Tooling-induced sampling bias}
In many organizations, sampling node pairs is mediated by security or management tooling, such as firewall policy managers and configuration databases.
These tools often expose some subsets of the network more readily than others, for example, privileging server-to-server communication over user endpoints or unmanaged devices.
Consequently, node pairs may be sampled from the set of entities visible to the tooling rather than uniformly from the underlying network, introducing visibility-driven bias.

\textbf{Configured vs enforced policy}
The metric is defined in terms of configured policies, but measurements in practice reflect observed enforcement.
The two may differ due to routing constraints, implicit deny rules, or configuration drift from recorded network security policy.

\textbf{Organizational and administrative constraints}
Beyond technical considerations, organizational factors may restrict which node pairs can be evaluated.
Access controls, change management processes, or business unit boundaries may limit the ability to probe certain communication paths.
As a result, practical sampling procedures may be partially randomized and partially constrained by administrative policy.

\textbf{Induced noise}
A critical operational detail regarding the proposed connectivity test suite is the potential for active probing to trigger security alerts within Intrusion Detection Systems (IDS)~\cite{dube2023faulty} or Security Information and Event Management (SIEM)~\cite{gonzalez2021security} platforms.
To mitigate potential disruption to security operations, measurements should ideally be scheduled during maintenance windows or coordinated with security teams to suppress expected alerts.
This allows for a controlled environment where the resulting alerts can be monitored and appropriately suppressed from standard escalation paths.

Taken together, these considerations highlight that sampling node pairs in real networks is an engineering challenge rather than a purely statistical one.
The segmentedness metric remains informative under such constraints when tracking relative changes over time, particularly when large-scale policy modifications (e.g., zero trust deployment, merger integration) produce measurable shifts in overall permissiveness that exceed the bias introduced by sampling constraints.

\section {Limitations} \label{sec:limitation}

The preceding section discussed operational challenges in deploying the measurement procedure.
This section addresses methodological limitations of the metric and estimator themselves.

\textbf{Binary connectivity model} The metric treats communication as binary (permitted/denied) without capturing degrees of restriction.
For example, connections limited to specific ports or protocols are treated identically to unrestricted connections. Future work could incorporate connection constraints into a weighted edge model.

\textbf{Static measurement} The metric captures policy-permitted connectivity at a point in time and does not reflect dynamic access control mechanisms, temporal restrictions, or context-dependent policies.
Networks using adaptive security models may require repeated measurements under different conditions.

\textbf{Symmetric edge assumption} As noted in Section~\ref{sec:network-model}, the undirected edge model assumes policy symmetry.
Networks with significant asymmetric policies (e.g., client-server architectures with one-way initiation requirements) may not be fully represented.

\textbf{Topology independence} While structural neutrality is a design goal (Section~\ref{sec:properties}), it means the metric does not capture important security-relevant properties such as hub-and-spoke vulnerabilities, critical chokepoints, or segmentation hierarchy.

\textbf{Sampling limitations} The statistical guarantees assume uniform random sampling and independent connectivity tests.
Non-uniform sampling strategies---such as oversampling from specific subnets or sampling geographically clustered nodes---could introduce correlation between samples and bias estimates.
Organizations should ensure sampling is truly random across the entire endpoint population.
See Section~\ref{sec:sampling} for a detailed discussion of practical sampling challenges in operational environments.

\textbf{Perimeter vs. internal} The metric focuses on internal network segmentation and does not characterize perimeter security, Internet connectivity, or external threat exposure.

\section {Conclusions} \label{sec:conclusion}

We introduce segmentedness as the first statistically principled scalar metric for network isolation, enabling practitioners to quantify (with explicit confidence intervals) what has previously required only qualitative assessment.
By formalizing segmentedness through edge density, we provide practitioners with an interpretable measure of how permissive network communication policies are.

The metric satisfies key properties for practical use: it is normalized for cross-network comparison, scales to networks of arbitrary size, and can be estimated efficiently through randomized sampling with explicit statistical guarantees.
Our validation demonstrates that estimates remain unbiased and well-behaved across different networks.

The main contribution of this paper is to provide a quantitative measure that practitioners can deploy immediately without specialized infrastructure or machine learning expertise.
Using standard network utilities and basic statistical knowledge, organizations can measure segmentedness with fewer than 100 sampled node pairs and track trends through quarterly measurements, validate security architecture initiatives, and establish empirical baselines: regardless of network size.

As organizations continue adopting zero trust architectures and micro-segmentation strategies, quantitative metrics like segmentedness will become essential tools for measuring progress, demonstrating compliance, and making data-driven security decisions.

\appendices

\section {Graph-based Measures' Comparison} \label{app:comparison}

Table~\ref{tab:structural-correlation} shows all three graph-based measures from Section~\ref{sec:graph-based} for the ten densest networks in an enterprise networking dataset~\cite{madani2022dataset,snap_cisco_data}.
The absolute correlation coefficient calculated over this table between Flatness and Modularity is $0.568$; that between Flatness and the Fiedler Value is $0.454$.
The moderate correlation of Flatness with Modularity and the Fiedler Value suggests the three measures capture distinct aspects of networks (graphs).

Note that 9 of the 10 networks are disconnected: $g7$ is the exception.
Since the Fiedler Value of disconnected graphs is 0, we report the Fiedler value of the largest connected component in Table~\ref{tab:structural-correlation}.
Calculating absolute correlation between Flatness and the Fiedler Value of the full (mostly disconnected) networks yields $0.136\ (< 0.454)$.

\begin{table}[htbp]
\centering
\caption{Graph-measures comparison: Flatness, Modularity, and Fiedler Value (largest connected component) of the ten densest networks in the enterprise networking dataset.}
\label{tab:structural-correlation}
\footnotesize
\begin{tabular*}{\columnwidth}{@{\extracolsep{\fill}}lrrr@{}}
\toprule
\textbf{Graph} & \textbf{$F(G)$} & \textbf{Modularity} & \textbf{Fiedler Value} \\
\midrule
g1           &    0.102 &      0.077 &      0.167 \\
g11          &    0.073 &      0.380 &      0.963 \\
g19          &    0.041 &      0.358 &      0.610 \\
g21          &    0.034 &      0.310 &      0.105 \\
g7           &    0.017 &      0.629 &      0.133 \\
g18          &    0.012 &      0.292 &      0.078 \\
g16          &    0.011 &      0.573 &      0.040 \\
g17          &    0.006 &      0.627 &      0.146 \\
g14          &    0.005 &      0.792 &      0.011 \\
g8           &    0.002 &      0.181 &      0.381 \\
\bottomrule
\end{tabular*}
\end{table}

\section{Deployment Guide} \label{app:deployment}

\subsection*{Prerequisites}

Organizations require three components to measure network segmentedness:

\textbf{Asset Inventory.}
A list of networked endpoints with IP addresses, typically maintained through configuration management databases (CMDBs), asset management systems, or network discovery tools.

\textbf{Connectivity Testing Tools.}
Standard network utilities for testing communication between node pairs, including \texttt{ping} (ICMP), \texttt{nc}/\texttt{netcat} (TCP/UDP), \texttt{nmap}, or equivalent tools available on Unix-like systems or Windows.

\textbf{Statistical Computing.}
Basic capability to compute sample statistics and confidence intervals using Python, R, Excel, or similar tools.

\subsection*{Measurement Procedure}

\textbf{Step 1: Enumerate nodes.}
Export the list of $n$ network endpoints from the asset inventory system.
Each entry should include a unique identifier (IP address or hostname) and be representative of systems subject to segmentation policy.

\textbf{Step 2: Sample node pairs.}
Generate $M$ random unordered pairs of distinct nodes.
For a 95\% confidence interval with $\pm 0.1$ half-width, set $M = 97$.
Use uniform random sampling with replacement from the set of all possible pairs.

\textbf{Step 3: Test connectivity.}
For each sampled pair $(u, v)$, execute the connectivity test suite: ICMP echo request (\texttt{ping -c 1 -W 1 <target>}), TCP connection attempts on common ports (\texttt{nc -zv -w 1 <target> 80 443}), and UDP probes (\texttt{nc -zuv -w 1 <target> 53}).
Mark the pair as connected if any test succeeds.
Record the number of connected pairs $k$.

\textbf{Step 4: Compute estimate.}
Calculate the flatness estimate $\hat{F} = k/M$ and segmentedness estimate $\hat{S} = 1 - \hat{F}$.

\textbf{Step 5: Compute confidence interval.}
The 95\% Wald confidence interval is $[\hat{S} - 0.1, \hat{S} + 0.1]$.
For the special case where $k = 0$, use the Bayesian approach described in Section~\ref{sec:bayesian} to avoid a zero-width interval.

\subsection*{Operational Recommendations}

\textbf{Scheduling.}
Coordinate connectivity testing with security operations teams to suppress expected alerts in IDS and SIEM platforms.
Schedule measurements during maintenance windows when possible to minimize operational disruption.

\textbf{Sampling bias.}
Ensure the asset inventory is current and complete before sampling.
Stale or incomplete inventories may bias measurements toward long-lived or well-managed systems, potentially overestimating segmentedness.

\textbf{Repeatability.}
Use the same connectivity test suite across all measurements to ensure comparability over time.
Document the specific tests used and any protocol-specific exclusions applied during measurement.

\textbf{Interpretation.}
A single measurement establishes a baseline value for network segmentedness.
Tracking $\hat{S}$ over time (e.g., quarterly) reveals trends in policy permissiveness and helps identify unintended drift or the impact of architectural changes.

\bibliographystyle{IEEEtran}
\bibliography{segflat}

@book{national2025cyber,
  address =       {Washington, DC, USA},
  author =        {{National Academies of Sciences, Engineering, and
                   Medicine}},
  note =          {{ISBN: 978-0-309-73489-9}},
  publisher =     {National Academies Press},
  title =         {Cyber Hard Problems: Focused Steps Toward a Resilient
                   Digital Future},
  year =          {2025},
}

@techreport{nist800207,
  address =       {Gaithersburg, MD, USA},
  author =        {Rose, S. and Borchert, O. and Mitchell, S. and
                   Connelly, S.},
  institution =   {Nat. Inst. Standards Technol., U.S. Dept. Commerce},
  month =         aug,
  note =          {doi: \url{https://doi.org/10.6028/NIST.SP.800-207}},
  number =        {800-207},
  type =          {NIST Special Publication},
  title =         {Zero Trust Architecture},
  year =          {2020},
}

@techreport{cisa2023ztmm,
  address =       {Washington, DC, USA},
  author =        {{Cybersecurity and Infrastructure Security Agency}},
  institution =   {Cybersecurity Infrastruct. Security Agency (CISA),
                   U.S. Dept. Homeland Security},
  month =         apr,
  note =          {Version 2.0.
  \url{https://www.cisa.gov/resources-tools/resources/zero-trust-maturity-model}},
  title =         {Zero Trust Maturity Model Version 2.0},
  year =          {2023},
}

@techreport{nistcsf20,
  address =       {Gaithersburg, MD, USA},
  author =        {{National Institute of Standards and Technology}},
  institution =   {Nat. Inst. Standards Technol., U.S. Dept. Commerce},
  month =         feb,
  note =          {\url{https://www.nist.gov/cyberframework}},
  number =        {CSWP 29},
  title =         {The {NIST} Cybersecurity Framework ({CSF}) 2.0},
  year =          {2024},
}

@inproceedings{wagner2016towards,
  address =       {Lexington, MA, USA},
  author =        {Wagner, N. and {\c{S}}ahin, C. {\c{S}}. and
                   Winterrose, M. and Riordan, J. and Pena, J. and
                   Hanson, D. and Streilein, W. W.},
  booktitle =     {Proc. IEEE Symp. Ser. Comput. Intell. (SSCI)},
  note =          {doi: \url{https://doi.org/10.1109/SSCI.2016.7849908}},
  pages =         {1--10},
  publisher =     {IEEE},
  title =         {Towards automated cyber decision support: A case
                   study on network segmentation for security},
  year =          {2016},
}

@inproceedings{sabur2019s3,
  address =       {Beijing, China},
  author =        {Sabur, A. and Chowdhary, A. and Huang, D. and
                   Kang, M. and Kim, A. and Velazquez, A.},
  booktitle =     {Proc. 22nd Int. Symp. Res. Attacks, Intrusions
                   Defenses (RAID)},
  month =         sep,
  note =          {{ISBN: 978-1-939133-07-6}},
  pages =         {473--485},
  publisher =     {USENIX Assoc.},
  title =         {{S3}: A {DFW}-based scalable security state analysis
                   framework for large-scale data center networks},
  year =          {2019},
}

@inproceedings{osman2020transparent,
  author =        {Osman, A. and Wasicek, A. and K{\"o}psell, S. and
                   Strufe, T.},
  booktitle =     {Proc. 3rd USENIX Workshop Hot Topics Edge Comput.
                   (HotEdge)},
  month =         jun,
  note =          {{ISBN: 978-1-7138-1525-9}},
  pages =         {1--6},
  publisher =     {USENIX Assoc.},
  title =         {Transparent microsegmentation in smart home {IoT}
                   networks},
  year =          {2020},
}

@inproceedings{basta2022towards,
  author =        {Basta, N. and Ikram, M. and Kaafar, M. A. and
                   Walker, A.},
  booktitle =     {Proc. IEEE/IFIP Netw. Oper. Manag. Symp. (NOMS)},
  note =          {doi:
                   \url{https://doi.org/10.1109/NOMS54207.2022.9789888}},
  pages =         {1--7},
  publisher =     {IEEE},
  title =         {Towards a zero-trust micro-segmentation network
                   security strategy: An evaluation framework},
  year =          {2022},
}

@article{mhaskar2021formal,
  author =        {Mhaskar, N. and Alabbad, M. and Khedri, R.},
  journal =       {Comput. Secur.},
  note =          {doi:
                   \url{https://doi.org/10.1016/j.cose.2020.102162}},
  pages =         {102162},
  title =         {A formal approach to network segmentation},
  volume =        {103},
  year =          {2021},
}

@misc{tyagi2025measuring,
  author =        {Tyagi, S. and Murugesan, G.},
  howpublished =  {arXiv preprint},
  month =         aug,
  note =          {arXiv:2508.21005,
                   \url{https://arxiv.org/abs/2508.21005}},
  title =         {Measuring ransomware lateral movement susceptibility
                   via privilege-weighted adjacency matrix
                   exponentiation},
  year =          {2025},
}

@inproceedings{bredesen2025network,
  author =        {Bredesen, R. and Mujeye, S.},
  booktitle =     {Proc. 8th Int. Conf. Softw. Eng. Inf. Manag. (ICSIM)},
  note =          {doi: \url{https://doi.org/10.1145/3725899.3725920}},
  pages =         {137--141},
  publisher =     {ACM},
  title =         {Network segmentation security with the implementation
                   of threats},
  year =          {2025},
}

@inproceedings{mujib2020performance,
  author =        {Mujib, M. and Sari, R. F.},
  booktitle =     {Proc. 12th Int. Conf. Inf. Technol. Elect. Eng.
                   (ICITEE)},
  note =          {doi:
                       \url{https://doi.org/10.1109/ICITEE49829.2020.9271749}},
  pages =         {27--32},
  publisher =     {IEEE},
  title =         {Performance evaluation of data center network with
                   network micro-segmentation},
  year =          {2020},
}

@article{kotha2020network,
  author =        {Kotha, N. R.},
  journal =       {Turk. J. Comput. Math. Educ.},
  note =          {doi:
                   \url{https://doi.org/10.61841/turcomat.v11i3.14942}},
  number =        {3},
  pages =         {3023--3030},
  title =         {Network segmentation as a defense mechanism for
                   securing enterprise networks},
  volume =        {11},
  year =          {2020},
}

@inproceedings{yatagha2023security,
  address =       {Berlin, Germany},
  author =        {Yatagha, R. and Waedt, K. and Schindler, J. and
                   Kirdan, E.},
  booktitle =     {Proc. INFORMATIK 2023},
  note =          {doi: \url{https://doi.org/10.18420/inf2023_204}},
  pages =         {2051--2070},
  publisher =     {GI},
  title =         {Security challenges and best practices for resilient
                   {IIoT} networks: Network segmentation},
  year =          {2023},
}

@article{alofeishat2024build,
  author =        {Al-Ofeishat, H. A. and Alshorman, R.},
  journal =       {Int. J. Comput. Digit. Syst.},
  month =         sep,
  note =          {doi: \url{https://doi.org/10.12785/ijcds/1601111}},
  number =        {1},
  pages =         {1499--1508},
  title =         {Build a secure network using segmentation and
                   micro-segmentation techniques},
  volume =        {16},
  year =          {2024},
}

@inproceedings{mani2025securing,
  address =       {Philadelphia, PA, USA},
  author =        {Mani, S. K. and Hsieh, K. and Segarra, S. and
                   Chandra, R. and Zhou, Y. and Kandula, S.},
  booktitle =     {Proc. 22nd USENIX Symp. Netw. Syst. Des. Implement.
                   (NSDI)},
  month =         apr,
  note =          {doi: \url{https://doi.org/10.5555/3767955.3768010}},
  pages =         {253--268},
  publisher =     {USENIX Assoc.},
  title =         {Securing public cloud networks with efficient
                   role-based micro-segmentation},
  year =          {2025},
}

@article{dube2026taxonomy,
  author =        {Dube, R.},
  journal =       {IEEE Access},
  note =          {doi:
                   \url{https://doi.org/10.1109/ACCESS.2026.3658250}},
  pages =         {16921--16935},
  title =         {A taxonomy of segmentation in network security},
  volume =        {14},
  year =          {2026},
}

@book{wasserman1994social,
  address =       {Cambridge, U.K.},
  author =        {Wasserman, S. and Faust, K.},
  note =          {doi: \url{https://doi.org/10.1017/CBO9780511815478}},
  publisher =     {Cambridge Univ. Press},
  title =         {Social Network Analysis: Methods and Applications},
  year =          {1994},
}

@article{newman2004finding,
  author =        {Newman, M. E. J. and Girvan, M.},
  journal =       {Phys. Rev. E},
  note =          {doi:
                   \url{https://doi.org/10.1103/PhysRevE.69.026113}},
  number =        {2},
  pages =         {1--15},
  title =         {Finding and evaluating community structure in
                   networks},
  volume =        {69},
  year =          {2004},
}

@article{fiedler1973algebraic,
  author =        {Fiedler, M.},
  journal =       {Czechoslov. Math. J.},
  note =          {doi: \url{https://doi.org/10.21136/CMJ.1973.101168}},
  number =        {2},
  pages =         {298--305},
  title =         {Algebraic connectivity of graphs},
  volume =        {23},
  year =          {1973},
}

@book{feller1991introduction,
  author =        {Feller, W.},
  publisher =     {Wiley},
  title =         {An Introduction to Probability Theory and Its
                   Applications},
  note =          {{ISBN: 978-0-471-25709-7}},
  volume =        {2},
  year =          {1991},
}

@article{agresti1998approximate,
  author =        {Agresti, A. and Coull, B. A.},
  journal =       {Amer. Statist.},
  note =          {doi:
                   \url{https://doi.org/10.1080/00031305.1998.10480550}},
  number =        {2},
  pages =         {119--126},
  title =         {Approximate is better than ``exact'' for interval
                   estimation of binomial proportions},
  volume =        {52},
  year =          {1998},
}

@incollection{bollobas2011random,
  author =        {Bollob{\'a}s, B.},
  booktitle =     {Modern Graph Theory},
  note =          {doi: \url{https://doi.org/10.1007/978-1-4612-0619-4_7}},
  pages =         {215-–252},
  publisher =     {Springer},
  title =         {Random graphs},
  year =          {2011},
}

@article{holland1983stochastic,
  author =        {Holland, P. W. and Laskey, K. B. and Leinhardt, S.},
  journal =       {Soc. Netw.},
  note =          {doi:
                   \url{https://doi.org/10.1016/0378-8733(83)90021-7}},
  number =        {2},
  pages =         {109--137},
  title =         {Stochastic blockmodels: First steps},
  volume =        {5},
  year =          {1983},
}

@inproceedings{madani2022dataset,
  author =        {Madani, Omid and Averineni, Sai Ankith and
                   Gandham, Shashidhar},
  booktitle =     {Proc. ACM Int. Workshop Security Privacy Analytics
                   (IWSPA)},
  note =          {doi: \url{https://doi.org/10.1145/3510548.3519368}},
  pages =         {100--104},
  title =         {A dataset of networks of computing hosts},
  year =          {2022},
}

@misc{snap_cisco_data,
  author =        {Leskovec, Jure and Krevl, Andrej},
  howpublished =  {[Online]},
  note =          {Available:
  \url{https://snap.stanford.edu/data/cisco-networks.html}. Accessed: Feb. 1,
  2026},
  title =         {{SNAP} Datasets: {Cisco} Secure Workload Networks of
                   Computing Hosts},
  year =          {2021},
}

@book{gelman2013bayesian,
  address =       {New York, NY, USA},
  author =        {Gelman, A. and Carlin, J. B. and Stern, H. S. and
                   Dunson, D. B. and Vehtari, A. and Rubin, D. B.},
  edition =       {3rd},
  note =          {doi: \url{https://doi.org/10.1201/b16018}},
  publisher =     {Chapman and Hall/CRC},
  series =        {Texts in Statistical Science},
  title =         {Bayesian Data Analysis},
  year =          {2013},
}

@article{dube2023faulty,
  author =        {Dube, Rohit},
  journal =       {J. Comput. Virol. Hacking Techn.},
  note =          {doi:
                   \url{https://doi.org/10.1007/s11416-023-00509-7}},
  pages =         {203--211},
  title =         {Faulty use of the {CIC-IDS} 2017 dataset in
                   information security research},
  volume =        {20},
  year =          {2023},
}

@article{gonzalez2021security,
  author =        {Gonz{\'a}lez-Granadillo, Gustavo and
                   Gonz{\'a}lez-Zarzosa, Susana and Diaz, Rodrigo},
  journal =       {Sensors},
  note =          {doi: \url{https://doi.org/10.3390/s21144759}},
  number =        {14},
  pages =         {4759},
  title =         {Security information and event management ({SIEM}):
                   Analysis, trends, and usage in critical
                   infrastructures},
  volume =        {21},
  year =          {2021},
}

\end{document}